\documentclass[structabstract]{aa}  
\usepackage{graphicx}
\usepackage{txfonts}
\usepackage{textcomp}
\usepackage{natbib}
\begin{document}

\title{Massive star formation by accretion}
\subtitle{I. Disc accretion}

\author{L. Haemmerl\'e\inst{\ref{inst1},\ref{inst2}},
P. Eggenberger\inst{\ref{inst1}},
G. Meynet\inst{\ref{inst1}},
A. Maeder\inst{\ref{inst1}},
C. Charbonnel\inst{\ref{inst1}}}
\authorrunning{Haemmerl\'e et al.}

\institute{Observatoire de Gen\`{e}ve, Universit\'{e} de Gen\`{e}ve, chemin des Maillettes 51, CH-1290 Sauverny, Switzerland        \label{inst1}
\and Institute f\"{u}r Theoretische Astrophysik, Zentrum f\"{u}r Astronomie der Universit\"{a}t Heidelberg,
Albert-Ueberle-Str. 2, D-69120 Heidelberg, Germany      \label{inst2}}

\date{Received 17 August 2015 ; accepted 4 November 2015}

% \abstract{}{}{}{}{} 
% 5 {} token are mandatory
 
 \abstract
 % context heading (optional)
 % {} leave it empty if necessary  
{
Massive stars likely form by accretion and the evolutionary track of an accreting forming star corresponds to what is called the birthline in the HR diagram.
The shape of this birthline is quite sensitive to the evolution of the entropy in the accreting star.
}
 % aims heading (mandatory)
{
We first study the reasons why some birthlines published in  past years present different behaviours for a given accretion rate.
We then revisit the question of the accretion rate, which  allows us to understand the distribution
of the observed pre-main-sequence (pre-MS) stars in the Hertzsprung-Russell (HR) diagram.
Finally,  we identify the conditions needed to obtain a large inflation of the star along its pre-MS evolution
that may push the birthline towards the Hayashi line in the upper part of the HR diagram.
}
 % methods heading (mandatory)
{
We present new pre-MS models including accretion at various rates and for different initial structures of the accreting core.
We compare them with previously published equivalent models.
From the observed upper envelope of pre-MS stars in the HR diagram,
we deduce the accretion law that best matches the accretion history of most of the intermediate-mass stars.
}
 % results heading (mandatory)
{
In the numerical computation of the time derivative of the entropy,
some treatment leads to an artificial loss of entropy and thus reduces the inflation that the accreting star undergoes along the birthline.
In the case of cold disc accretion,
the existence of a significant swelling during the accretion phase, which leads to radii $\gtrsim100\,R_\odot$
and brings  the star back to the red part of the HR diagram, depends sensitively on the initial conditions.
For an accretion rate of $10^{-3}\,M_\odot\rm\,yr^{-1}$,
only models starting from a core with a significant radiative region evolve back to the red part of the HR diagram.
We also obtain  that, in order to reproduce the observed upper envelope of pre-MS stars in the HR diagram
with an accretion law deduced from the observed mass outflows in ultra-compact HII regions,
the fraction of the mass that is accreted onto the star
should represent a decreasing fraction of the mass outflows when the mass of the accreting object increases.
In other words, the accretion efficiency (mass effectively accreted onto the star with respect to the total in falling matter)
decreases when the mass of the star increases.
}
 % conclusions heading (optional), leave it empty if necessary 
 {}
 
   \keywords{Stars: formation -- Stars: evolution -- Accretion, accretion discs}
 
\maketitle
%
%________________________________________________________________

\section{Introduction}
\label{sec-intro}

Massive stars are rare, but their influence on our Universe is critical.
They give the main chemical and dynamical inputs to the interstellar medium.
Most of the heavy elements in our Universe were produced in the core of massive stars or during their deaths.
The mass-loss during their lifetimes, and more dramatically their deaths in supernovae,
feed the interstellar medium with these newly produced heavy elements.
Such outflows and explosions, as well as the stellar radiation itself,
are also thought to be the main mechanisms that trigger star formation.
Thus, massive stars drive the star formation in their host galaxies and make them evolve.
However, the formation of massive stars is currently the most poorly known part of their evolution.

Stars form in the gravitational collapse of cold dense clouds.
Since the collapse is highly non-homologous, hydrostatic equilibrium is first reached by a small central core
while the rest of the cloud is still collapsing, nearly in free-fall
(e.g. \citealt{mcnally1964}, \citealt{bodenheimer1968}, \citealt{larson1969,larson1972}).
As the collapse proceeds, the hydrostatic core evolves as a pre-main-sequence (pre-MS) star,
with a mass increasing with time.
This is the accretion scenario
(e.g. \citealt{shu1977}, \citealt{stahler1980a,stahler1980b,stahler1986a,stahler1986b}, \citealt{palla1990,palla1991,palla1992}).

In the case of spherical symmetry, it has been shown that this scenario is not valid for massive stars
($M\gtrsim40\,M_\odot$) owing to their strong radiation field (\citealt{kahn1974,yorke1977,wolfire1987}).
But we know now that an accretion flow with a disc geometry allows this limitation to be circumvented
(\citealt{nakano1989,yorke2002,krumholz2009,peters2010a,kuiper2010,kuiper2011}).
Hydrodynamic simulations for the accretion flow in massive proto-stellar clouds (e.g.~\citealt{peters2010a,kuiper2010}),
as well as observations of high-mass protostellar objects (e.g.~\citealt{fazal2008}),
suggest that the typical value of the accretion rate during the formation of massive stars is $\sim 10^{-3}\,M_\odot\rm\,yr^{-1}$.
This is in contrast with the case of low-mass stars, for which values of $\sim10^{-5}\,M_\odot\rm\,yr^{-1}$ are expected
(see e.g.~\citealt{stahler1980a,stahler1980b,stahler1983}).

The accretion scenario for massive star formation has been explored by many authors in a stellar evolution approach.
\cite{bernasconi1996a}, \cite{norberg2000}, \cite{behrend2001}, and \cite{haemmerle2013}
have computed stellar models with accretion using the Geneva Stellar Evolution code.
The accretion rates used in these works were increasing functions of time, i.e. of the actual stellar mass.
In the low-mass range ($M\lesssim2\,M_\odot$) values around $10^{-5}\,M_\odot\rm\,yr^{-1}$ were considered,
while in the high-mass range ($M\gtrsim10\,M_\odot$) the values were $\sim10^{-3}\,M_\odot\rm\,yr^{-1}$.

\cite{hosokawa2009} and \cite{hosokawa2010} have also computed stellar models with accretion for the formation of massive stars,
using their own code.
The accretion rates they used were constant in time (i.e. independent of the actual stellar mass),
and so already had high values ($\sim10^{-3}\,M_\odot\rm\,yr^{-1}$) in the low-mass range.
They found that when a star accretes at such high rates in the intermediate-mass range
($2\,M_\odot\lesssim M\lesssim10\,M_\odot$), it goes through a rapid swelling phase
where the stellar radius can increase by more than one order of magnitude,
leading to values as high as several $100\,R_\odot$ depending on the physical conditions.
They also found  that above a critical accretion rate of $\simeq2-3\times10^{-3}\,M_\odot\rm\,yr^{-1}$,
a second swelling occurs before the star can contract to the zero-age-main-sequence (ZAMS),
leading to stellar radii near $1000\,R_\odot$.
In this case, at a mass around $\sim50\,M_\odot$, steady accretion is no longer possible
owing to the high radiation pressure that approaches the Eddington limit.

Since the works listed above were published, the treatment of accretion in the Geneva Stellar Evolution code
has been improved (\citealt{haemmerle2014}).
In the original version of the code, an inconsistency in the treatment of the time derivatives
was responsible for an artificial loss of entropy during the accretion phase.
This inconsistency has been removed in the new version of the code.
In the present paper, we show how these improvements modify pre-MS evolution with accretion.
The paper is organized as follows.
In Sect.~\ref{sec-code}, we describe the improvements brought to the code.
In Sect.~\ref{sec-cst}, we present models computed with the new version of the code using constant accretion rates,
and we compare the results with those obtained with the original version of the code
and with those of other authors.
In Sect.~\ref{sec-ch}, we do the same with the time-dependent accretion rate used by \cite{behrend2001} and \cite{haemmerle2013}.
The conclusions are summarized in Sect.~\ref{sec-outro}.

\section{Improvements in the code}
\label{sec-code}

A full description of the original treatment of accretion in the Geneva Stellar Evolution code
has already been given  in the previous papers (\citealt{bernasconi1996a,haemmerle2013}).
Here we focus on the improvements implemented since the last paper.
A full description of the treatment of accretion in the new version of the code is available in \cite{haemmerle2014}.
The treatment for the burning of light elements is also described in \cite{haemmerle2013},
and a general description of the other physical ingredients in the code is available in \cite{eggenberger2008}.

The new improvements are related to the time variation of the thermodynamic quantities.
Among the four equations of structure, the equation of energy conservation is the only one that involves time,
so this is the equation that makes the star evolve.
The equation is obtained from the energetic balance of a given mass element $dm$.
A fraction of the energy generated by nuclear reactions is transformed into heat, the rest is radiated,
\begin{equation}
\int\epsilon_n\,dm\,=\int{dq\over dt}\,dm+\int{\nabla\vec{F}\over\rho}dm
,\end{equation}
where $\epsilon_n$ is the energy generation rate by nuclear reactions, $q$ the heat by unit of mass,
$t$ the time, $\vec{F}$ the flux, and $\rho$ the mass density.
Since this equality is satisfied for any mass element, we have
\begin{equation}
\epsilon_n={dq\over dt}+{\nabla\vec{F}\over\rho}
=T\,{ds\over dt}+{dL_r\over dM_r}
\label{eq-e1},\end{equation}
where $s$ is the specific entropy, $T$ the temperature, $L_r$ the luminosity in the shell of radius $r$,
and $M_r$ the total mass contained in the interior of this shell.
We emphasize that
\begin{equation}
{d\over dt}=\left.{d\over dt}\right\vert_{M_r}
\end{equation}
since we started from the energetic balance of a given mass element (Lagrangian coordinates).

An alternative treatment is to use the relative mass coordinate
\begin{equation}
\mu={M_r\over M}
,\end{equation}
which is not a Lagrangian coordinate in the case of accretion since $M$ increases with time.
With this coordinate, Eq.~(\ref{eq-e1}) becomes
\begin{equation}
\epsilon_n=T\left.{ds\over dt}\right\vert_\mu-{\mu\dot M\over M}\cdot T{ds\over d\mu}+{1\over M}{dL_r\over d\mu}
\label{eq-e2}.\end{equation}
Equation.~(\ref{eq-e2}) contains one more term than Eq.~(\ref{eq-e1}),
which is proportional to the accretion rate and the entropy gradient.
Since the entropy increases outwards and $\dot M>0$ during accretion, this additional term is negative.
This procedure has been  used  by e.g. \cite{stahler1980a,stahler1980b} and \cite{hosokawa2009}.
These two methods are physically equivalent.

In our code, we keep the Lagrangian formulation using Eq.~(\ref{eq-e1}).
However, in the original version, we treated the differential $ds$ as a time variation at fixed $\mu$ instead of fixed $M_r$.
In the constant mass case, such a treatment is fully consistent since a fixed $\mu$ also corresponds to a fixed $M_r$.
However, in the case of accretion, it leads to an artificial loss of entropy.
This inconsistency is physically equivalent to the neglect of the additional term in Eq.~(\ref{eq-e2}).
If we neglect this negative term, this is equivalent to assuming that  $\left.{ds\over dt}\right\vert_{M_r} = \left.{ds\over dt}\right\vert_\mu$, while
actually $\left.{ds\over dt}\right\vert_{M_r} =\left.{ds\over dt}\right\vert_\mu-{\mu\dot M\over M}\cdot T{ds\over d\mu}$.
Thus, we see that the neglect of the additional term implies a value for the time derivative of the entropy that is too large or in other words,
an artificial loss of entropy.
This inconsistency has been removed in the new version of the code: we still use Eq.~(\ref{eq-e1}),
but the differential $ds$ that appears in this equation is now treated consistently as a time variation at constant $M_r$.

For the accretion of entropy (i.e. the computation of $ds/dt$ in the layer newly accreted at each time step),
we make the assumption that the entropy of the material that is accreted
is the same as the entropy of the stellar surface before the material is accreted.
This is the assumption of \textit{cold disc accretion} (see e.g.~\citealt{palla1992,hosokawa2010,haemmerle2013}).
It corresponds to the case of an accretion flow with a thin disc geometry,
allowing any entropy excess to be radiated away in the polar directions before it is advected in the stellar interior.
Thus, we do not add any contribution from the gravitational energy liberated by the material when it is accreted,
and the entropy profiles are built continuously.
This assumption is generally considered as the lower limit for the accretion of entropy (e.g.~\citealt{hosokawa2010}).
The corresponding boundary conditions are the usual photospheric boundary conditions.

We use the Schwarzschild criterion for convection without overshooting.
The chemical composition is solar ($Z=0.014$) with the abundances of \cite{asplund2005} and \cite{cunha2006}.
For deuterium, we take a mass fraction of $5\times10^{-5}$
as in \cite{bernasconi1996a}, \cite{norberg2000}, \cite{behrend2001}, and \cite{haemmerle2013}.

\section{Models with constant accretion rates}
\label{sec-cst}

\subsection{Case with $\dot M=10^{-4}\,M_\odot\rm\,yr^{-1}$}
\label{sec-cst-4}

\begin{figure}
\includegraphics[width=0.49\textwidth]{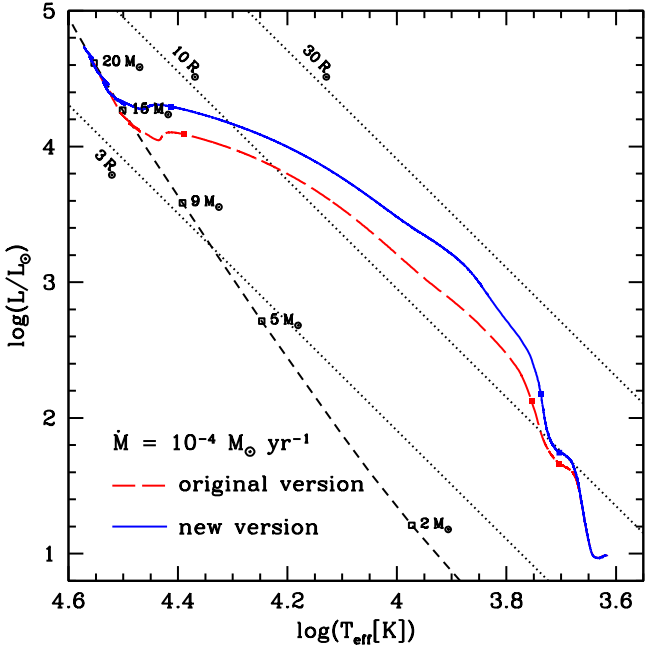}
\caption{Evolutionary tracks of two models having the same initial conditions (Eq.~\ref{eq-ini4})
accreting at the same rate ($10^{-4}\,M_\odot\rm\,yr^{-1}$), but computed with the two different versions of the code.
The black dotted straight lines are iso-radius for the indicated values of $R$,
and the black short-dashed line is the ZAMS of \cite{ekstroem2012} with a few masses indicated.
From bottom to top, the filled squares on each track are the apparition of the radiative core,
the end of the convective envelope, and the apparition of the convective core 
(see Fig.~\ref{fig-st4}).}
\label{fig-hr4}
\end{figure}

\begin{figure}
\includegraphics[width=0.49\textwidth]{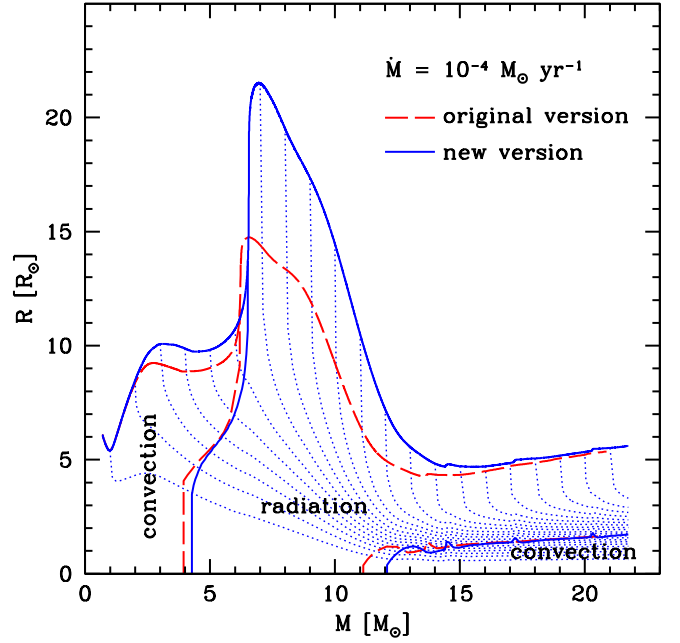}
\caption{Evolution of the stellar radius and the internal structure for the same models as in Fig.~\ref{fig-hr4}.
The horizontal axis corresponds to the stellar mass, which is a time coordinate in the case of accretion.
In each case, the upper line indicates the stellar radius, the other continuous and long-dashed lines
indicate the boundary between radiation and convection.
The thin blue dotted lines are iso-mass of 1, 2, 3, ..., $21\,M_\odot$
for the model computed with the new version of the code.}
\label{fig-st4}
\end{figure}

In order to illustrate the effect of the improvements described in Sect.~\ref{sec-code},
we first consider a pre-MS star accreting at a constant rate of
\begin{equation}
\dot M=10^{-4}\,M_\odot\rm\,yr^{-1}
\label{eq-dm4}\end{equation}
For the initial model, we consider
\begin{equation}
M=0.7\,M_\odot    \qquad      L=9.57\,L_\odot    \qquad    T_{\rm eff}=4130\,K
\label{eq-ini4}\end{equation}
This initial model has a radius of $6.06\,R_\odot$ and is fully convective.
Starting from this initial model, we compute two models accreting at the rate of Eq.~(\ref{eq-dm4}),
one with the original version of the code, the other with the new version,
including the improvements described in Sect.~\ref{sec-code}.
The evolutionary tracks and the evolution of the radii and the internal structures of both models
are shown in Figs.~\ref{fig-hr4} and \ref{fig-st4}.

In the first stages of the evolution (low-mass range, $M\lesssim2\,M_\odot$), both models are identical.
In convective regions, the entropy gradient is flat,
so that the time variation of entropy at fixed $\mu$ or at fixed $M_r$ is the same.

The initial value of the central temperature is too low for a significant D-burning,
and the star takes its energy from gravitational contraction.
At $M\simeq1\,M_\odot$, the central temperature reaches $\sim10^6K$ and D-burning becomes significant.
Owing to its high sensitivity in temperature ($\epsilon_{\rm D}\sim T^{11.7}$, see e.g.~\citealt{maeder2009}),
D-burning has a thermostatic effect, leading to a central temperature $T_c$ that is nearly constant until the end of central burning.
Since $T_c\sim M/R$, it produces an increase in the stellar radius, nearly linearly with respect to the stellar mass (see Fig.~\ref{fig-st4}).
The deuterium mass fraction decreases in the whole star, and goes to 0 at $M\simeq2-3\,M_\odot$.
D-burning becomes too weak for the star and gravitational contraction starts again.

It is at this point that both models start to diverge
with a smaller radius for the original model than for the new model.
But while the star is still fully convective, the differences remain relatively modest.
The small differences that appear are due to the non-adiabaticity of convection in the external regions.

Since central D-burning is now negligible, $T_c$ can increase again, and the opacity decreases.
At $M\simeq4\,M_\odot$, a radiative core appears, earlier in the original model ($M=3.8\,M_\odot$)
than in the new model ($M=4.3\,M_\odot$).
The huge decrease in opacity in the deep regions leads to the so-called \textit{luminosity wave}
(e.g.~\citealt{larson1972,hosokawa2010}):
the low opacity in the radiative core imposes a flux that is too strong for the convective envelope, in which opacity is still high.
The convective envelope absorbs the flux, so that the luminosity profile has a maximum near the boundary between these two regions.
As the temperature increases in the star, the radiative core grows in mass and the maximum in luminosity moves outwards.
The energy absorbed by the convective envelope leads to an increase in the stellar radius,
which becomes particularly strong as the convective envelope recedes.

It is at this point that the differences between the two models become really significant.
While the original model reaches only $14.7\,R_\odot$, the new one reaches $21.5\,R_\odot$, which is 46\% higher.
Because of the strong entropy gradient that develops in the radiative regions,
a $ds$ at a given $\mu$ or at a given $M_r$ are significantly different.

When the whole star becomes radiative, the radius decreases again, and the star contracts towards the ZAMS.
Around $11-12\,M_\odot$, the triggering of the CN cycle produces a convective core,
and the energy released by the nuclear reactions replaces the gravitational contraction as the main energy source of the star.
Contraction slows down until the stellar radius reaches a minimum, which can be seen as the ZAMS
at masses of $13.6\,M_\odot$ and $15.6\,M_\odot$ for the original and the new model, respectively.
At this point, the differences between the two models become small again:
for the original model the minimum radius is $4.3\,R_\odot$, while it is $4.7\,R_\odot$ for the new model (only 9\% higher).
Then, as accretion goes on the stellar radius increases nearly linearly with the stellar mass,
following the ZAMS (see Fig.~\ref{fig-hr4}).
We stopped the computation when the central mass fraction of hydrogen decreased by 0.3\% with respect to its initial value.

\quad

To summarize, the differences between the two models are as follows:
\begin{itemize}
\item the new model evolves more slowly than the original;
\item the new model reaches higher radii than the original, mainly during the swelling phase.
\end{itemize}
The other differences remain modest.

\quad

Why does the neglect of the additional term have the strongest effect during the swelling?
Because the swelling occurs in the intermediate-mass range, when the star is fully radiative,
having thus a strong entropy gradient.
In this case, the artificial loss of entropy in the original model becomes significant.
Using the homology relations $P\sim M^2/R^4$ and $T\sim M/R$ the specific entropy reads
\begin{equation}
s\sim-\ln{P\over T^{5/2}}\sim\ln(MR^3)
\label{eq-s}\end{equation}
(\citealt{hosokawa2009}).
It follows that, for a given mass, the smaller the entropy is, the smaller the radius.
As a consequence, the artificial loss of entropy leads to artificially small radii, softening the swelling.
Furthermore, since the gravitational contraction -- which governs pre-MS evolution -- corresponds to a decrease of the entropy,
the artificial loss of entropy accelerates pre-MS evolution.

\subsection{Case with $\dot M=10^{-3}\,M_\odot\rm\,yr^{-1}$}
\label{sec-cst-3}

\begin{figure}
\includegraphics[width=0.49\textwidth]{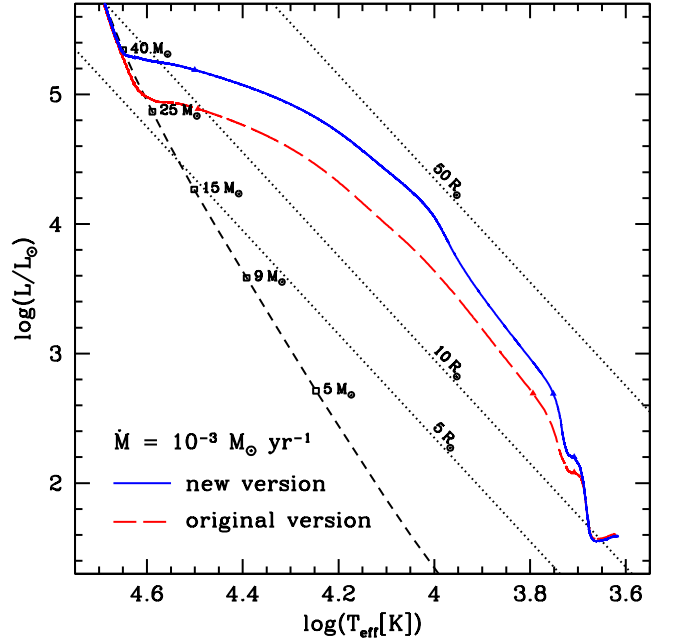}
\caption{Same as Fig.~\ref{fig-hr4}, but for an accretion rate of $10^{-3}\,M_\odot\rm\,yr^{-1}$
and the initial model CV1.}
\label{fig-hr3}
\end{figure}

\begin{figure}
\includegraphics[width=0.49\textwidth]{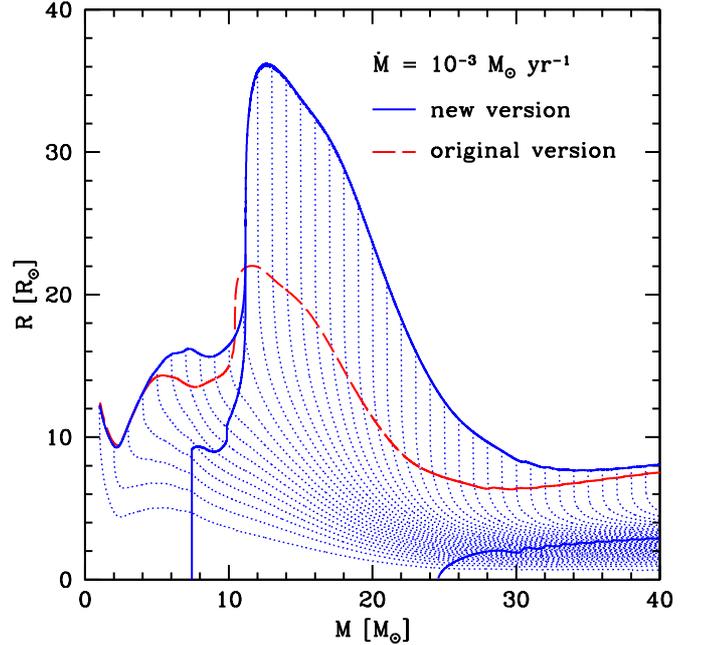}
\caption{Same as Fig.~\ref{fig-st4}, but for an accretion rate of $10^{-3}\,M_\odot\rm\,yr^{-1}$
and the initial model CV1.
(The internal structure is shown only for the new model.)}
\label{fig-st3}
\end{figure}

The typical value of the accretion rate for massive star formation
is higher than the $10^{-4}\,M_\odot\rm\,yr^{-1}$ considered in Sect.~\ref{sec-cst-4}.
In the present section, we make the same comparison as in Sect.~\ref{sec-cst-4}, but using the rate
\begin{equation}
\dot M=10^{-3}\,M_\odot\rm\,yr^{-1}
\label{eq-dm3}.\end{equation}
For the initial model, we first consider
\begin{equation}
M=1.0\,M_\odot    \qquad    L=37.9\,L_\odot    \qquad     T_{\rm eff}=4130\,K
\label{eq-ini3}\end{equation}
labelled CV1, which has a radius of $R=12.1\,R_\odot$ and is again fully convective\footnote{
   This initial model is not the same as in Sect.~\ref{sec-cst-4} (Eq.~\ref{eq-ini4}).
   Owing to convergence difficulties, it was not possible until now to compute a model accreting at the rate of Eq.~(\ref{eq-dm3})
   starting from the initial model given by Eq.~(\ref{eq-ini4}).}.
The evolutionary tracks and the evolution of the radii and the internal structures
of models starting from CV1 accreting at the rate of Eq.~(\ref{eq-dm3})
computed with both versions of the code are shown in Figs.~\ref{fig-hr3} and \ref{fig-st3}.

The result is qualitatively similar to the case with $10^{-4}\,M_\odot\rm\,yr^{-1}$.
The original and the new model are identical until the end of central D-burning.
Then the new model has larger radii and evolves more slowly than the original model.
Again, the main difference is the magnitude of the swelling:
while the maximum value of the stellar radius is $22.0\,R_\odot$ for the original model,
it is $36.2\,R_\odot$ for the new model, which makes a relative difference of 65\%.
It illustrates the fact that the artificial loss of entropy is more important when the accretion rate is higher.
For a given time-step $dt$ and a given value of $\mu$, the difference in $M_r$ between the model of age $t$
and the model of age $t+dt$ increases with the accretion rate since
\begin{equation}
dM_r=\mu\,dM=\mu\,\dot Mdt
.\end{equation}

\quad

In order to test the dependence of our results on the initial model,
we compute two additional models with the new version of the code
at the same rate of $10^{-3}\,M_\odot\rm\,yr^{-1}$, but using two different initial models, labelled CV2 and RC.

The CV2 model is given by
\begin{equation}
M=2.0\,M_\odot    \qquad    L=123\,L_\odot    \qquad     T_{\rm eff}=4270\,K
\label{eq-inicv}.\end{equation}
This model has a radius of $20.4\,R_\odot$ and is fully convective, as is CV1.
Both models are fully convective; CV2 differs from CV1 only in terms of global properties, and not in terms of internal structure.
We note, however, that the CV2 model has a larger radius than the model of the same mass obtained by accretion starting from CV1.
With CV1, the stellar radius reached when the mass is equal to $2\,M_\odot$ is only $9.4\,R_\odot$,
so about a factor of two smaller than the radius of CV2.
In agreement with Eq.~(\ref{eq-s}), the larger radius of the CV2 model corresponds to a higher total amount of entropy.

Finally, the RC model is given by
\begin{equation}
M=2.0\,M_\odot    \qquad    L=4.32\,L_\odot    \qquad     T_{\rm eff}=5230\,K
\label{eq-inirc}.\end{equation}
It has a radius of $2.53\,R_\odot$ and is thus much more compact than CV1 and CV2.
Moreover, in contrast to all the previous cases, the initial RC model is not fully convective,
but is made of a radiative core ($M_r<1.53\,M_\odot$) and a convective envelope.
We note that for both CV2 and RC, we used a mass of $2\,M_\odot$, which is higher than for CV1.
Because of numerical difficulties, it is easier to compute accreting models
with the structure of RC at $M=2\,M_\odot$ than at $M=1\,M_\odot$.
In order to allow a direct comparison, we used the same mass for CV2.

\begin{figure}
\includegraphics[width=0.49\textwidth]{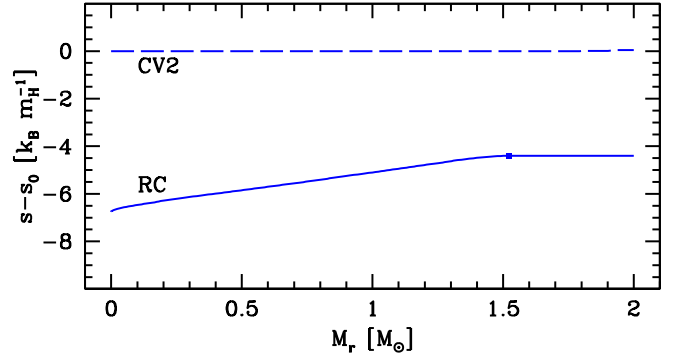}
\caption{Entropy profiles of the initial models CV2 and RC.
The square along the RC profile indicates the boundary between the radiative core and the convective envelope.}
\label{fig-dsini}
\end{figure}

In terms of thermodynamic quantities, the main difference between the initial models CV2 and RC is the entropy profile.
Both entropy profiles are shown in Fig.~\ref{fig-dsini}.
First, we see that the global entropy is lower in the RC case than in the CV2 case, which was expected from Eq.~(\ref{eq-s}).
Moreover, in contrast to the entropy profile of CV2, which is flat (corresponding to a fully convective structure with adiabatic convection),
the entropy profile of RC increases outwards (except in the convective envelope where it is flat).
In the radiative core, the entropy profile increases nearly linearly with the Lagrangian coordinate with a gradient given by
\begin{equation}
\left.{ds\over dM_r}\right\vert_{\rm rad}\simeq1.3\ k_B\,m_H^{-1}\,M_\odot^{-1}
\label{eq-ds}.\end{equation}
In other words, the RC model differs from the CV2
model not only in terms of global properties, but also (and more importantly) in terms of internal structure.

Another difference between these initial models is the deuterium abundance.
The low central entropy of the RC model also corresponds to a high central temperature, which exceeds the temperature for D-burning:
$T_c=6.1\times10^6\,K$, instead of $7.7\times10^5\,K$ in the case CV2.
Indeed, Eq.~(\ref{eq-s}) and $T_c\sim M/R$ give $s\sim\ln(M^4/T_c^3)$, so that for a given mass,
the lower is the entropy, the higher is the central temperature.
With such a high temperature, we expect all the deuterium contained originally in the material of this ``initial'' model
to be destroyed in the previous stages of the evolution, and we take $X_2=0$ in the initial model RC.
However, we keep the usual mass fraction ($X_2=5\times10^{-5}$, Sect.~\ref{sec-code}) in the accreted material, as in the CV cases.

\begin{figure}
\includegraphics[width=0.49\textwidth]{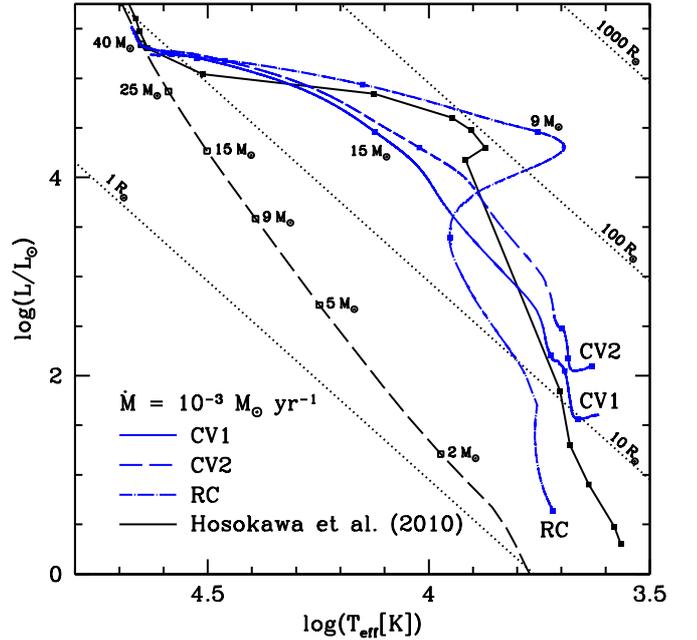}
\caption{Evolutionary tracks of the models computed from the initial models CV1, CV2 and RC
at $\dot M=10^{-3}\,M_\odot\rm\,yr^{-1}$, using the new version of the code.
For comparison, a series of points from the model MD3-D of \cite{hosokawa2010} is indicated with connected black squares.
The black dashed curve is the ZAMS of \cite{ekstroem2012} with a few masses indicated,
and the black dotted straight lines are iso-radius of 1, 10, 100 and $1000\,R_\odot$.
The blue squares along the tracks correspond to the same masses as those indicated along the ZAMS.
(Notice that, on the vertical axis, $L$ is the intrinsic stellar luminosity, without any contribution from the accretion luminosity.)}
\label{fig-hrini}
\end{figure}

The evolutionary tracks obtained from these two new initial models are shown in Fig.~\ref{fig-hrini},
together with the track obtained from CV1 (which is the same as the blue track in Fig.~\ref{fig-hr3}).
The evolution obtained from CV2 is similar to the CV1 case.
The star is initially fully convective, until a radiative core appears, here at a mass of $8.8\,M_\odot$.
Then the convective envelope recedes and disappears,
while the luminosity wave described in Sect.~\ref{sec-cst-4} produces the swelling of the star.
The radius reaches its maximum value of $48.6\,R_\odot$ at a mass of $12.6\,M_\odot$,
before the star contracts towards the ZAMS (reached at $M\sim30\,M_\odot$).
This example shows that a change in the total amount of entropy in the initial model,
without any difference in the internal distribution of the entropy,
does not result in a significant difference in the stellar evolution during the accretion phase.
For instance, while the initial radius of the CV2 model is more than a factor of 2 larger than that of CV1,
the maximum value reached by the radius during the swelling is increased by only 34\%.¥

The evolution corresponding to the RC model shows significant differences compared to the CV cases (i.e. CV1 and CV2).
In the case of RC with its early radiative core, the evolution starts with the luminosity wave and the swelling.
Owing to the low stellar luminosity, the external layers radiate away the entropy that they absorb from the central regions
with less efficiency than in the CV cases: the Kelvin-Helmholtz time is longer.
As a consequence, the accretion produces a stronger swelling,
leading to radii higher than $100\,R_\odot$ already in the intermediate-mass range.
In particular, the radius reaches its maximum value of $204\,R_\odot$ at a mass of $7.5\,M_\odot$.
At this stage, the luminosity is sufficiently high to radiate the energy that was stored in the envelope
and that was responsible for the swelling of the star.
The accreting star decreases in radius and the birthline joins the ZAMS (reached again at $M\sim30\,M_\odot$).

We see that in this case the maximum value reached by the radius during the accretion phase
is  nearly one order of magnitude higher than in the CV cases.
It shows that a change in the internal distribution of entropy in the initial model
results in significant differences in the stellar evolution during the accretion phase.

\quad

The rate considered in the present section (Eq.~\ref{eq-dm3}) is the same as the one used by \cite{hosokawa2010}.
The model of these authors reached radii of $\sim100\,R_\odot$ during the swelling,
while our CV models reach only $\sim40\,R_\odot$. 
However, we saw with the previous examples that the maximum value reached by the radius during the swelling
depends on the initial model with a high sensitivity.
When we use the initial model RC,
we obtain values for the maximum radius that are higher than those of \cite{hosokawa2010} by a factor of 2.

In the fiducial model of \cite{hosokawa2010}, the initial entropy profile is not flat. 
It increases outwards, linearly with the Lagrangian coordinate:
\begin{equation}
s=s_c+{k_B\over m_H}\cdot{M_r\over M}
\label{eq-s-hoso}.\end{equation}
This initial model is thus fully radiative, instead of fully convective. 
Their initial mass ($0.1\,M_\odot$) is also lower by one order of magnitude than in our case, so that a direct comparison is not possible.
However, if we use the same assumption of Eq.~(\ref{eq-s-hoso}) for a $2\,M_\odot$ star,
corresponding to the mass of our initial models CV2 and RC, we obtain an entropy gradient of
\begin{equation}
\left.{ds\over dM_r}\right\vert_{\rm H10}=0.5\,[k_B\,m_H^{-1}\,M_\odot^{-1}]
\label{eq-ds-hoso}.\end{equation}
Thus we can see the fiducial model of \cite{hosokawa2010} as an intermediate case
between our models CV2, with a flat entropy profile, and RC, with a gradient given essentially by Eq.~(\ref{eq-ds}),
\begin{equation}
\left.{ds\over dM_r}\right\vert_{\rm CV2}<\left.{ds\over dM_r}\right\vert_{\rm H10}<\left.{ds\over dM_r}\right\vert_{\rm RC}
\label{eq-ds3};\end{equation}
however, the second inequality holds only in the radiative core of the model RC, i.e. for $M_r\lesssim1.5\,M_\odot$.
In Fig.~\ref{fig-hrini}, we added a series of points from the fiducial model of \cite{hosokawa2010} (their model labelled MD3-D).
We see that  the HR track of this model also corresponds  to an intermediate case between our tracks CV2 and RC,
which is consistent with Eq.~(\ref{eq-ds3}).

Moreover, \cite{hosokawa2010} tested the dependence of their results on the choice of the initial model
by considering different initial entropy profiles.
In particular, they used an initial model with a flat entropy profile (labelled MD3-cv), which is thus fully convective.
In this case,  during the swelling they obtained a maximum radius of only $30\,R_\odot$  (see their Fig.~9),
which is even lower than the value we obtain in our models with a flat initial entropy profile (CV cases).
We  note that the total amount of entropy in their initial model is also lower than in our cases.
Taking into account these differences in the initial conditions, it appears that
for given physical conditions our models are in good agreement with those of \cite{hosokawa2010}.

It follows from the above examples that the maximum radius reached by a given star
and the possibility for it to evolve towards the red part of the HR diagram
depends sensitively on the initial conditions, namely on the initial entropy profile,
which reflects the physics of the pre-stellar collapse, and is currently not known.

\section{Models with the Churchwell-Henning accretion rate}
\label{sec-ch}

\subsection{Accretion rate}
\label{sec-ch-dm}

As mentioned in Sect.~\ref{sec-intro}, pre-MS models with accretion for the formation of massive stars
have been computed using an accretion rate increasing with time,
instead of the constant accretion rates described in Sect.~\ref{sec-cst}.
In particular, \cite{behrend2001} and \cite{haemmerle2013} used an accretion rate based on the Churchwell-Henning relation.
This relation, established by \cite{churchwell1999} and confirmed by \cite{henning2000},
is an empirical correlation between the mass-loss rates through outflows $\dot M_{\rm out}$ in ultra-compact HII regions
and the bolometric luminosity $L$ of the corresponding central sources.
A polynomial fit gives (\citealt{behrend2001})
\begin{equation}\begin{array}{l}
\log\,(\dot M_{\rm out}[M_\odot{\rm\,yr^{-1}}])=-5.28+0.752\,\log\,(L[L_\odot])
\\\qquad\qquad\qquad\qquad\qquad\qquad\qquad-0.0278\,\log^2(L[L_\odot]).
\end{array}\label{eq-ch}\end{equation}
The correlation covers 6 orders of magnitude in luminosity, from $\sim L_\odot$ to $\sim10^6L_\odot$,
corresponding to ZAMS luminosities on the whole intermediate- and high-mass range.

Such mass outflows are expected to be produced at the inner boundary of accretion discs
by mechanisms such as magnetic field or radiative heating in a way that is still poorly understood (\citealt{maeder2009}).
From the material coming from the disc, a fraction $f$ is accreted by the central star,
the rest ($1-f$) is rejected in the polar directions, producing the outflows.
If we know the value of $f$, the Churchwell-Henning relation provides an accretion rate,
given as a function of the bolometric luminosity of the accreting object.
Indeed, since $\dot M=f\,\dot M_{\rm disc}$ and $\dot M_{\rm out}=(1-f)\,\dot M_{\rm disc}$, Eq.~(\ref{eq-ch}) gives
\begin{equation}\begin{array}{l}
\log\,({1-f\over f}\dot M[M_\odot{\rm\,yr^{-1}}])=-5.28+0.752\,\log\,(L[L_\odot])
\\\qquad\qquad\qquad\qquad\qquad\qquad\qquad-0.0278\,\log^2(L[L_\odot]).
\end{array}\label{eq-dmch}\end{equation}

This procedure was used first by \cite{behrend2001}.
Since the mechanisms producing the outflows are not yet understood, the value of $f$ is not known.
With the assumption that $f$ is constant (independent of the mass of the accreting object),
\cite{behrend2001} computed birthlines\footnote{
   The \textit{birthline} is the set of the points on the HR diagram where the stars of various masses become optically visible.
   With the assumption that the accretion history $\dot M(M)$ is unique, which can be partially justified with  Newton's theorem,
   the birthline is also the evolutionary track of each individual accreting star.
   If this assumption is correct, such a track has to reproduce the upper envelope of observed pre-MS objects.
   Without this assumption, the birthline can be seen as the upper envelope of each individual track.
   See \cite{maeder2009} for more details.}
with different values of $f$.
By comparing the birthlines with the location of observed Herbig Ae/Be stars on the HR diagram,
they found that the best fit was given by the value $f=1/3$ (i.e. $(1-f)/f=2$).
With this value of $f$, they obtained a birthline which is in agreement with observations on the whole stellar-mass range.
With a constant $f$, Eq.~(\ref{eq-dmch}) gives an accretion rate that increases with $L$ and thus with $M$.
We note that, since the relation between $M$ and $L$ is given by the stellar models,
the accretion history $\dot M(M)$ also depends on the stellar models.
With $f=1/3$, the values of the accretion rate obtained by \cite{behrend2001} were $\sim10^{-5}\,M_\odot\rm\,yr^{-1}$ in the low-mass range,
$\sim10^{-4}\,M_\odot\rm\,yr^{-1}$ in the intermediate-mass range, and $\sim10^{-3}\,M_\odot\rm\,yr^{-1}$ in the high-mass range.
With this accretion rate, they were able to produce stars of $85\,M_\odot$ still close to the ZAMS.

However, these results were obtained with the version of the code that overestimated the losses of entropy (see Sect.~\ref{sec-code}).
It is therefore interesting to reconsider this question with the new version.
This is the aim of the present section.
We compute birthlines with the new version of the code, using the accretion rate given by Eq.~(\ref{eq-dmch})
and various values of $f$. For $L$, we consider the intrinsic stellar luminosity without any contribution from the accretion luminosity,
as in \cite{behrend2001}.

\subsection{Model with $f=1/3$}
\label{sec-ch-f3}

\begin{figure}
\includegraphics[width=0.49\textwidth]{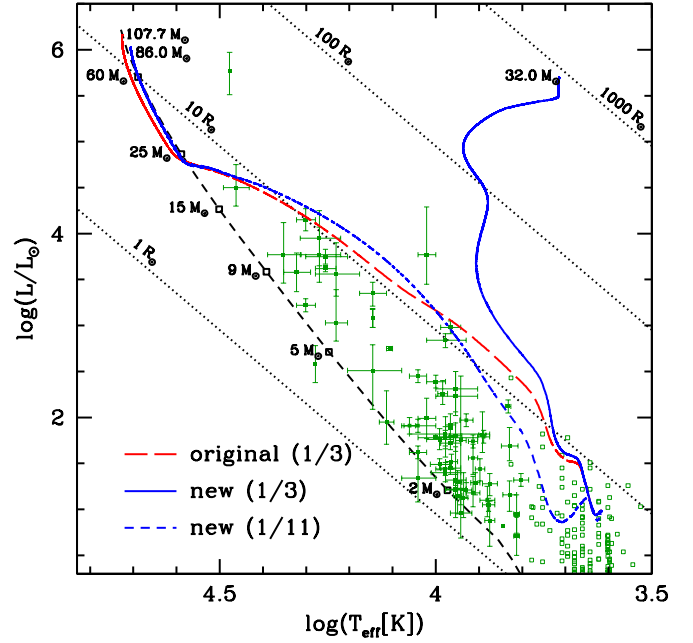}
\caption{Birthlines computed with the Churchwell-Henning accretion rate (Eq.~\ref{eq-dmch})
and the same initial model (Eq.~\ref{eq-inich}), but with different versions of the code (labelled ``original'' and ``new'')
and different values of the fraction $f$ that appears in Eq.~(\ref{eq-dmch}), $f=1/3$ and $1/11$.
The green dots indicate observations of pre-MS stars
(Herbig Ae/Be stars by \cite{alecian2013a}, filled squares with error bars; T Tauri stars by \cite{cohen1979}, empty squares).
The black short-dashed line is the ZAMS of \cite{ekstroem2012}, with a few masses indicated,
and the black dotted straight lines are iso-radius of indicated values.
The final mass of each model is indicated at the end of the corresponding track.}
\label{fig-hrch}
\end{figure}

\begin{figure}
\includegraphics[width=0.49\textwidth]{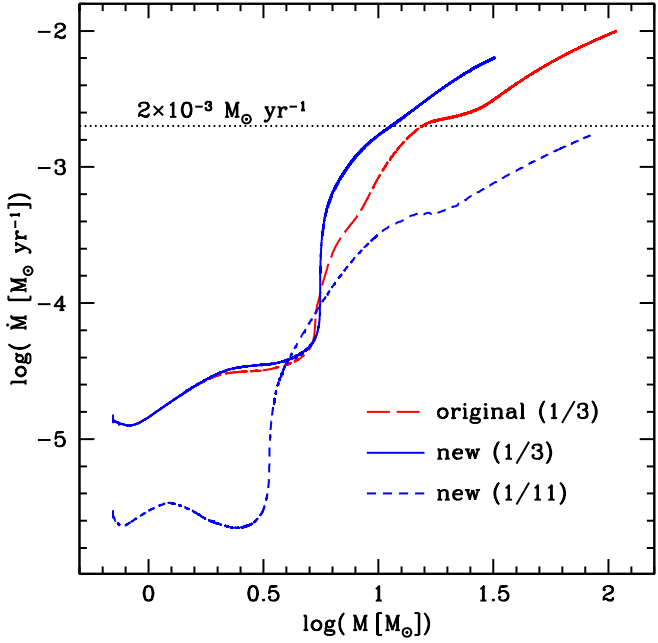}
\caption{Accretion histories of the same models as in Fig.~\ref{fig-hrch}.
The black dotted horizontal line indicates the critical rate of \cite{hosokawa2009} and \cite{hosokawa2010}.}
\label{fig-dmch}
\end{figure}

We first consider the Churchwell-Henning accretion rate (Eq.~\ref{eq-dmch}) with the same value $f=1/3$ as used by \cite{behrend2001}.
For the initial model, we take
\begin{equation}
M=0.7\,M_\odot    \qquad    L=13.2\,L_\odot    \qquad     T_{\rm eff}=4102\,K
\label{eq-inich}\end{equation}
and a solar composition, as in Sect.~\ref{sec-cst}.
This initial model has a radius of $7.2\,R_\odot$ and is again fully convective.
As in Sect.~\ref{sec-cst-4} and \ref{sec-cst-3}, we compute birthlines starting with the same initial model (Eq.~\ref{eq-inich}),
accreting at the same rate\footnote{
   i.e. the same $\dot M(L)$, but not the same $\dot M(M)$, as mentioned in Sect.~\ref{sec-ch-dm}.}
(Eq.~\ref{eq-dmch}; $f=1/3$), but with the two different versions of the code (Sect.~\ref{sec-code}).
The evolutionary tracks and the accretion histories of this models are shown in Figs.~\ref{fig-hrch} and \ref{fig-dmch}
(long-dashed red line and solid blue line).
The birthline obtained in this way with the original version of the code (red dashed),
is the same as that  of \cite{behrend2001}.
We see that this birthline closely corresponds to the empirical upper envelope
of the pre-MS intermediate-mass stars observed by \cite{alecian2013a} and the pre-MS low-mass stars observed by \cite{cohen1979},
indicated by the green dots.

\begin{figure}
\includegraphics[width=0.49\textwidth]{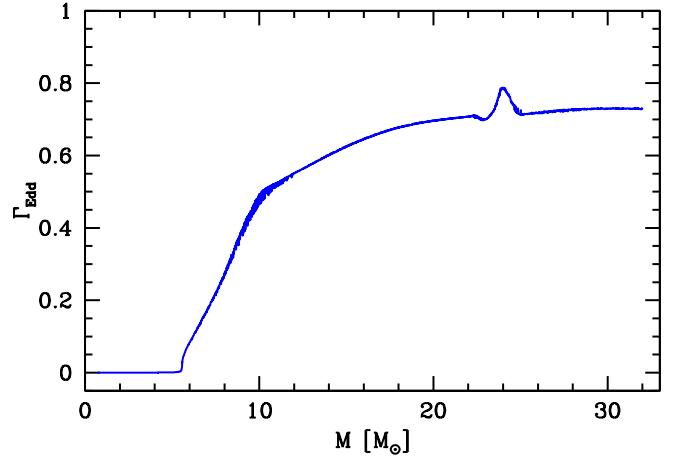}
\caption{Eddington factor as a function of the stellar mass, for the model ``new (1/3)'' of Fig.~\ref{fig-hrch},
i.e. of the model computed with the accretion rate of Eq.~(\ref{eq-dmch}) with $f=1/3$
using the new version of the code (Sect.~\ref{sec-code}), starting from the initial model given by Eq.~(\ref{eq-inich}).}
\label{fig-edch}
\end{figure}

As in the case with a constant accretion rate (Sect.~\ref{sec-cst}), both models remain essentially identical in the low-mass range,
while the star is fully convective.
But again, when the star enters the intermediate-mass range, becomes radiative, and starts its swelling, the two models diverge;
the new model has  a higher radius and luminosity than the original model (Fig.~\ref{fig-hrch}).
However, in this case, the dependence of the accretion rate on the stellar luminosity produces a positive feedback:
the rapid increase in luminosity in the new model leads to a rapid increase in its accretion rate (see  Fig.~\ref{fig-dmch}).
In this model, the accretion rate reaches $10^{-3}\,M_\odot\rm\,yr^{-1}$ already in the intermediate-mass range (at $M=7.3\,M_\odot$),
while in the original model it is reached only in the high-mass range (at $M=10.6\,M_\odot$).
As a consequence, the swelling is enhanced in the new model compared to the original one,
leading to radii as high as $100\,R_\odot$ (Fig.~\ref{fig-hrch}) at a mass of $10.2\,M_\odot$.
At such a mass, which corresponds roughly to the beginning of the high-mass range, the swelling slows down,
but before the star can contract again, the positive feedback between the stellar luminosity and the accretion rate
leads to the critical value of Hosokawa (Fig.~\ref{fig-dmch}, at a mass of $11.5\,M_\odot$).
At this point, the stellar luminosity exceeds 50\% of the Eddington luminosity (Fig.~\ref{fig-edch}),
and the second swelling occurs (see Sect.~\ref{sec-intro} or \citealt{hosokawa2009,hosokawa2010} for more details).
As we can see in Fig.~\ref{fig-hrch}, during this second swelling, the star goes back to the red
until it reaches the boundary of the forbidden region of Hayashi, as in the case of a red giant.
Then, as the swelling goes on, the star moves upwards on the HR diagram, along a vertical line, similar to the red giant branch
(i.e. it moves up the Hayashi line).
We stopped the computation when the stellar mass was $32\,M_\odot$ with a radius approaching $1000\,R_\odot$
and a luminosity close to 75\% of the Eddington luminosity (Fig.~\ref{fig-edch}).
If we stop accretion at this point, the star simply contracts towards the ZAMS in a Kelvin-Helmholtz time
($45\,000\rm\,yr$ for such a mass), as in the canonical scenario of pre-MS at constant mass.
It shows that the canonical scenario corresponds to the limit of the accretion scenario for a high accretion rate.

\begin{figure}
\includegraphics[width=0.49\textwidth]{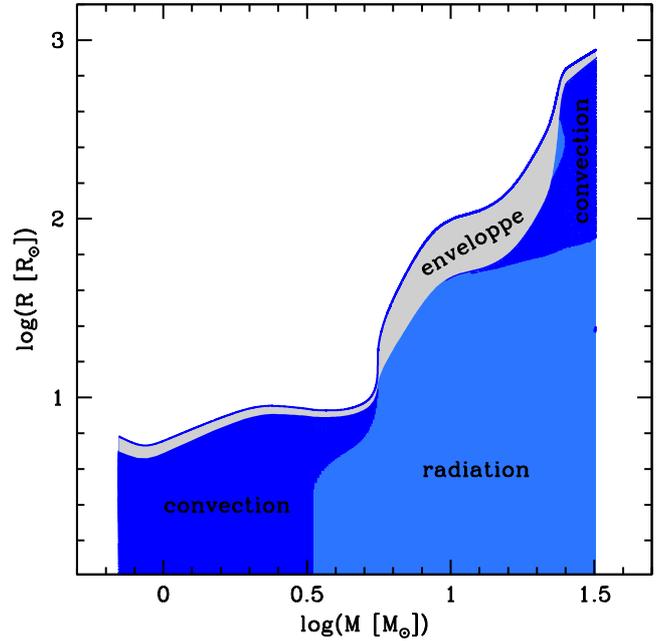}
\caption{Evolution of the radius (upper blue curve) and the internal structure of the model ``new (1/3)'' of Fig.~\ref{fig-hrch},
i.e. of the model computed with the accretion rate of Eq.~(\ref{eq-dmch}) with $f=1/3$
using the new version of the code (Sect.~\ref{sec-code}), starting from the initial model given by Eq.~(\ref{eq-inich}).
The dark blue areas are convective zones, the light blue areas are radiative regions,
and the grey area is the envelope (see footnote \ref{note-fitm}).}
\label{fig-stch}
\end{figure}

The internal structure of this model is shown in Fig.~\ref{fig-stch}.\footnote{
   The grey area is the envelope, a region that is not treated as the stellar interior by the code.
   In this region, where convection is not adiabatic and the ionization is not complete,
   we make the simplifying assumption that $L_r=L$.\label{note-fitm}}
Until the end of the first swelling, the internal structure is similar to the cases of constant accretion rates described in Sect.~\ref{sec-cst}.
But when the second swelling occurs, the convective envelope comes back,
as in the case of a red giant or in the initial model of a $\sim30\,M_\odot$ star in the canonical scenario.

The new birthline computed with $f=1/3$, if  representative of the birthline for the majority of stars,
would predict a much more extended distribution than the one observed for the pre-MS stars. 
In the next section we discuss a birthline obtained with a lower value of $f$, which produces an upper envelope
more compatible with the observations.

\subsection{Model with $f=1/11$}
\label{sec-ch-f11}

We computed a third birthline with the new version of the code, the same initial model,
and the Churchwell-Henning accretion rate, but with a value of $f=1/11$ (i.e. $(1-f)/f=10$).
The evolutionary track of this birthline is shown in Fig.~\ref{fig-hrch} (blue dashed line),
and the corresponding accretion history is in Fig.~\ref{fig-dmch}.

As we can see, thanks to the low value of $f$, the accretion rate of this new birthline is lower than in the other two cases
by nearly one order of magnitude.
This new birthline diverges rapidly from the other two already in the low-mass range, but when the star enters the intermediate-mass range,
the swelling brings  the new birthline back to the original one (see Fig.~\ref{fig-hrch}).
Then, the two birthlines remain close to each other until the end of the computation at masses above $85\,M_\odot$.
In particular, with this new birthline, we can also reach such high masses still close to the ZAMS.

\subsection{Discussion}
\label{sec-ch-bla}

As we can see by looking at the two new birthlines in Sects.~\ref{sec-ch-f3} and \ref{sec-ch-f11} (blue tracks in Fig.~\ref{fig-hrch}),
if we want to reproduce the original birthline, which is in agreement with the observations in the whole stellar-mass range,
it is necessary to consider a value of $f$ that decreases with the stellar mass.
With the original value of $f=1/3$, we still obtain a satisfactory result in the low-mass range,
but in the intermediate- and high-mass range, such a high value of $f$ leads to luminosities
that are too high compared to those of Herbig Ae/Be stars.
If such an accretion history was realistic, we would expect to observe massive pre-MS stars
contracting towards the ZAMS at such high luminosities, which is not the case.\footnote{\ 
   However, the Kelvin-Helmholtz time of such massive stars is very short,
   which could also explain the absence of stars in this region of the HR diagram.}
Inversely, if we consider the value $f=1/11$,  in the low-mass range we obtain an accretion rate that is much lower
than the typical rate of $10^{-5}\,M_\odot\rm\,yr^{-1}$  expected for such masses (e.g.~\citealt{stahler1980a,stahler1980b}).
With this accretion rate, we obtain luminosities that are too low compared to the observed T Tauri stars,
as can be seen  in Fig.~\ref{fig-hrch}.
But in the intermediate- and high-mass range, this new value of $f$ leads to a birthline that is in good agreement with the observations,
fitting the upper envelope of Herbig Ae/Be stars and allowing us to produce stars of masses above $80\,M_\odot$ still close to the ZAMS.
Interestingly, this value is also in good agreement with the observational estimations
of \cite{churchwell1999} for B-type stars, which corresponds to the intermediate-mass range.
As a consequence, our results suggest that the efficiency of accretion from a disc decreases with the stellar mass:
the higher the stellar mass, the larger the fraction of the material coming from the disc that is rejected through the bipolar outflows.

The assumption of a unique accretion history is certainly an approximation,
and we expect the real accretion histories of stars to depend on the final mass they are destined to reach.
In this sense, each of our birthlines has to be seen as an upper envelope of the evolutionary tracks
of each individual accreting star, corresponding to the main accretion phase (\citealt{maeder2009}).
Even in this case, however,    the value of $f$ during the main accretion phase
has to be lower for stars of intermediate- and high-mass than for low-mass stars
in order to reproduce the upper envelope of observed pre-MS stars.

We note that when we compute the accretion rate from the luminosity with Eq.~(\ref{eq-dmch}),
we do not include the accretion luminosity $L_{\rm accr}$.
This additional contribution would increase the accretion rate, leading to a higher birthline on the HR diagram.
In order to keep the birthline in agreement with the observations, we would need to reduce the value of $f$.
Thus the value $f=1/11$ has to be seen as an upper limit for the intermediate- and high-mass range.
Moreover, $L_{\rm accr}$ represents a larger fraction of the total luminosity $L_{\rm tot}=L+L_{\rm accr}$
in the low-mass range than in the intermediate- and high-mass ranges.
In the low-mass range, $L_{\rm accr}\simeq90\%L_{\rm tot}$ typically,
while in the intermediate- and high-mass ranges $L_{\rm accr}\lesssim50\%L_{\rm tot}$.
As a consequence, taking into account the accretion luminosity would reduce the decrease of $f$ as a function of $M$.

Another issue is the dependence of these results on the initial model.
 In Sect.~\ref{sec-cst-3} we showed that for a constant accretion rate of $10^{-3}\,M_\odot\rm\,yr^{-1}$
the choice of the initial model has a critical impact on the evolution.
However, in the case of the Churchwell-Henning accretion law for $f=1/11$,
the accretion rate remains much lower than $10^{-3}\,M_\odot\rm\,yr^{-1}$ during the whole pre-MS phase (Fig.~\ref{fig-dmch}),
and it turns out that the shape of this birthline is nearly unchanged when we use the initial model RC of Sect.~\ref{sec-cst-3}
instead of the initial model given by Eq.~(\ref{eq-inich}).
In other words, the choice of the initial model has no significant influence on the results of the present section.

\subsection{Main sequence evolution and isochrones}
\label{sec-ch-ms}

\begin{figure}
\includegraphics[width=0.49\textwidth]{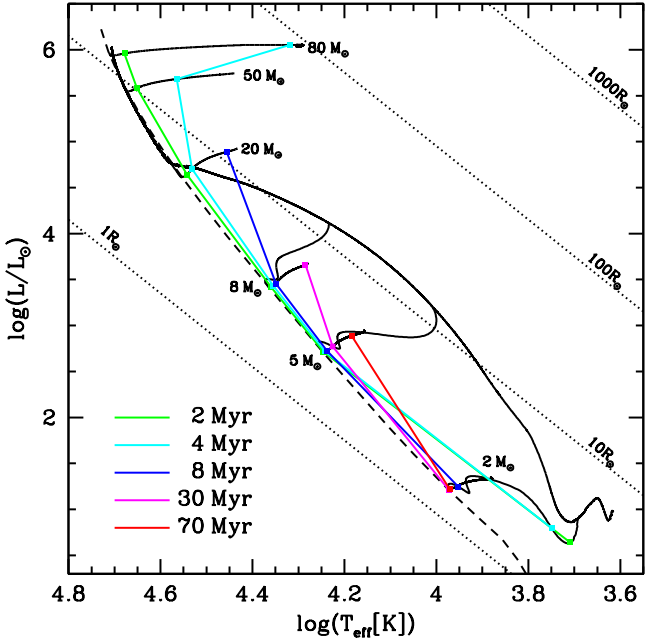}
\caption{HR tracks of the birthline ``new(1/11)'' (Sect.~\ref{sec-ch-f11}), and the subsequent contractions and MS tracks
for $M=2$, 5, 8, 20, 50, and $80\,M_\odot$ (solid black lines).
Several isochrones are indicated with coloured solid lines.
The iso-radius of 1, 10, 100, and $1000\,R_\odot$ are indicated with black dotted lines
and the ZAMS of \cite{ekstroem2012} is shown with a black dashed line.}
\label{fig-iso}
\end{figure}

Usually, MS models are computed starting from the ZAMS with initial conditions obtained from canonical pre-MS at constant mass.
However, \cite{bernasconi1996a} obtained in their models that the internal structure of a MS star
is not the same in the accretion scenario as in the canonical scenario.
They found that for a given mass and  central mass fraction of hydrogen, a MS star formed by accretion
has a smaller convective core than a MS star coming from a canonical pre-MS,
leading thus to a shorter MS lifetime.\footnote{\ 
   This result was obtained with the original version of the Geneva Stellar Evolution code, see Sect.~\ref{sec-code}.}
In order to see how the pre-MS scenario described in the previous sections modifies the structure and the evolution
of MS stars compared to canonical MS models, we computed contractions\footnote{\ 
   We call \textit{contractions}, the tracks followed by forming stars once they have reached their final masses and
   evolve at constant mass between the birthline and the ZAMS.}
at constant masses from the birthline described in Sect.~\ref{sec-ch-f11} (Churchwell-Henning accretion rate with $f=1/11$),
and the subsequent MS evolution.
We obtain that the internal structure on the ZAMS and the evolution on the MS are identical to those of canonical MS stars.
The HR tracks of the contractions for $M=2$, 5, 8, 20, 50, and $80\,M_\odot$
and the subsequent MS tracks are shown in Fig.~\ref{fig-iso}.
We also added several isochrones of 2, 4, 8, 30, and 70 Myr.
A more complete grid with isochrones will be available in a forthcoming paper.

\section{Conclusions and perspectives}
\label{sec-outro}

We showed that the previous versions of the code contained an inconsistency that lead to an artificial loss of entropy
in the case of an accreting star dominated by the radiative transport.
As a result of this artificial loss of entropy, we underestimated the magnitude of the swelling that occurs
when an intermediate-mass star accretes at a high rate ($\sim10^{-3}\,M_\odot\rm\,yr^{-1}$).
With the new version of the code, we obtain results that are in good agreement with those of other authors,
provided that we compare models with the same physical conditions.
In this context, we showed that the existence of a significant swelling that brings  the star back towards the red part of the HR diagram,
as described by other authors, depends critically on the initial conditions.

We looked at the effects of these improvements on the accretion history given by the Churchwell-Henning relation and
considered various values of $f$, the fraction of the material coming from the disc which is effectively accreted by the star.
Using the original value of $f=1/3$, we found that the first swelling of the star leads to a radius of $100\,R_\odot$;
before the star can contract again, a second swelling occurs, which leads the star to the red part of the HR diagram
until the boundary of the forbidden region of Hayashi with radii near $1000\,R_\odot$ and a new convective envelope.
We showed that instead, with a fraction $f=1/11$, the birthline remains close to the original birthline of \cite{behrend2001}
in the intermediate- and high-mass range.
This new value of $f$ is in good agreement with the estimations of \cite{churchwell1999} and provides a good fit
to the observed upper envelope of  pre-MS stars in the HR diagram for luminosities higher than about 2.5.
At lower luminosities, a value of $f$ equal to 1/3 gives a better fit.
Therefore,  in order to reproduce these observations with the new version of the code,
we conclude that it is necessary to consider a fraction $f$ that decreases with the stellar mass,
which suggests that, when the stellar mass increases,
the mass that is accreted represents a still smaller fraction of the mass that is in the outflows.

In the next paper in this series we will add the effects of rotation.
Since pre-MS models simultaneously including the effects of accretion and rotation and computed with the original version of the code
have been published (\citealt{haemmerle2013}), we will describe how the new improvements modify the results in the case of rotation.

\begin{acknowledgements}
Part of this work was supported by the Swiss National Science Foundation.
\end{acknowledgements}

\bibliographystyle{aa}
\bibliography{bibliotheque}

\begin{thebibliography}{39}
\expandafter\ifx\csname natexlab\endcsname\relax\def\natexlab#1{#1}\fi

\bibitem[{{Alecian} {et~al.}(2013){Alecian}, {Wade}, {Catala}, {Grunhut},
  {Landstreet}, {Bagnulo}, {B{\"o}hm}, {Folsom}, {Marsden}, \&
  {Waite}}]{alecian2013a}
{Alecian}, E., {Wade}, G.~A., {Catala}, C., {et~al.} 2013, \mnras, 429, 1001

\bibitem[{{Asplund} {et~al.}(2005){Asplund}, {Grevesse}, \&
  {Sauval}}]{asplund2005}
{Asplund}, M., {Grevesse}, N., \& {Sauval}, A.~J. 2005, in Astronomical Society
  of the Pacific Conference Series, Vol. 336, Cosmic Abundances as Records of
  Stellar Evolution and Nucleosynthesis, ed. T.~G. {Barnes}, III \& F.~N.
  {Bash}, 25

\bibitem[{{Behrend} \& {Maeder}(2001)}]{behrend2001}
{Behrend}, R. \& {Maeder}, A. 2001, \aap, 373, 190

\bibitem[{{Bernasconi} \& {Maeder}(1996)}]{bernasconi1996a}
{Bernasconi}, P.~A. \& {Maeder}, A. 1996, \aap, 307, 829

\bibitem[{{Bodenheimer} \& {Sweigart}(1968)}]{bodenheimer1968}
{Bodenheimer}, P. \& {Sweigart}, A. 1968, \apj, 152, 515

\bibitem[{{Churchwell}(1999)}]{churchwell1999}
{Churchwell}, E. 1999, in NATO ASIC Proc. 540: The Origin of Stars and
  Planetary Systems, ed. C.~J. {Lada} \& N.~D. {Kylafis}, 515

\bibitem[{{Cohen} \& {Kuhi}(1979)}]{cohen1979}
{Cohen}, M. \& {Kuhi}, L.~V. 1979, \apjs, 41, 743

\bibitem[{{Cunha} {et~al.}(2006){Cunha}, {Hubeny}, \& {Lanz}}]{cunha2006}
{Cunha}, K., {Hubeny}, I., \& {Lanz}, T. 2006, \apjl, 647, L143

\bibitem[{{Eggenberger} {et~al.}(2008){Eggenberger}, {Meynet}, {Maeder},
  {Hirschi}, {Charbonnel}, {Talon}, \& {Ekstr{\"o}m}}]{eggenberger2008}
{Eggenberger}, P., {Meynet}, G., {Maeder}, A., {et~al.} 2008, \apss, 316, 43

\bibitem[{{Ekstr{\"o}m} {et~al.}(2012){Ekstr{\"o}m}, {Georgy}, {Eggenberger},
  {Meynet}, {Mowlavi}, {Wyttenbach}, {Granada}, {Decressin}, {Hirschi},
  {Frischknecht}, {Charbonnel}, \& {Maeder}}]{ekstroem2012}
{Ekstr{\"o}m}, S., {Georgy}, C., {Eggenberger}, P., {et~al.} 2012, \aap, 537,
  A146

\bibitem[{{Fazal} {et~al.}(2008){Fazal}, {Sridharan}, {Qiu}, {Robitaille},
  {Whitney}, \& {Zhang}}]{fazal2008}
{Fazal}, F.~M., {Sridharan}, T.~K., {Qiu}, K., {et~al.} 2008, \apjl, 688, L41

\bibitem[{{Haemmerl{\'e}}(2014)}]{haemmerle2014}
{Haemmerl{\'e}}, L. 2014, PhD thesis, Universit{\'e} de Gen{\`e}ve

\bibitem[{{Haemmerl{\'e}} {et~al.}(2013){Haemmerl{\'e}}, {Eggenberger},
  {Meynet}, {Maeder}, \& {Charbonnel}}]{haemmerle2013}
{Haemmerl{\'e}}, L., {Eggenberger}, P., {Meynet}, G., {Maeder}, A., \&
  {Charbonnel}, C. 2013, \aap, 557, A112

\bibitem[{{Henning} {et~al.}(2000){Henning}, {Schreyer}, {Launhardt}, \&
  {Burkert}}]{henning2000}
{Henning}, T., {Schreyer}, K., {Launhardt}, R., \& {Burkert}, A. 2000, \aap,
  353, 211

\bibitem[{{Hosokawa} \& {Omukai}(2009)}]{hosokawa2009}
{Hosokawa}, T. \& {Omukai}, K. 2009, \apj, 691, 823

\bibitem[{{Hosokawa} {et~al.}(2010){Hosokawa}, {Yorke}, \&
  {Omukai}}]{hosokawa2010}
{Hosokawa}, T., {Yorke}, H.~W., \& {Omukai}, K. 2010, \apj, 721, 478

\bibitem[{{Kahn}(1974)}]{kahn1974}
{Kahn}, F.~D. 1974, \aap, 37, 149

\bibitem[{{Krumholz} {et~al.}(2009){Krumholz}, {Klein}, {McKee}, {Offner}, \&
  {Cunningham}}]{krumholz2009}
{Krumholz}, M.~R., {Klein}, R.~I., {McKee}, C.~F., {Offner}, S.~S.~R., \&
  {Cunningham}, A.~J. 2009, Science, 323, 754

\bibitem[{{Kuiper} {et~al.}(2010){Kuiper}, {Klahr}, {Beuther}, \&
  {Henning}}]{kuiper2010}
{Kuiper}, R., {Klahr}, H., {Beuther}, H., \& {Henning}, T. 2010, \apj, 722,
  1556

\bibitem[{{Kuiper} {et~al.}(2011){Kuiper}, {Klahr}, {Beuther}, \&
  {Henning}}]{kuiper2011}
{Kuiper}, R., {Klahr}, H., {Beuther}, H., \& {Henning}, T. 2011, \apj, 732, 20

\bibitem[{{Larson}(1969)}]{larson1969}
{Larson}, R.~B. 1969, \mnras, 145, 271

\bibitem[{{Larson}(1972)}]{larson1972}
{Larson}, R.~B. 1972, \mnras, 157, 121

\bibitem[{{Maeder}(2009)}]{maeder2009}
{Maeder}, A. 2009, {Physics, Formation and Evolution of Rotating Stars}
  (Springer)

\bibitem[{{McNally}(1964)}]{mcnally1964}
{McNally}, D. 1964, \apj, 140, 1088

\bibitem[{{Nakano}(1989)}]{nakano1989}
{Nakano}, T. 1989, \apj, 345, 464

\bibitem[{{Norberg} \& {Maeder}(2000)}]{norberg2000}
{Norberg}, P. \& {Maeder}, A. 2000, \aap, 359, 1025

\bibitem[{{Palla} \& {Stahler}(1990)}]{palla1990}
{Palla}, F. \& {Stahler}, S.~W. 1990, \apjl, 360, L47

\bibitem[{{Palla} \& {Stahler}(1991)}]{palla1991}
{Palla}, F. \& {Stahler}, S.~W. 1991, \apj, 375, 288

\bibitem[{{Palla} \& {Stahler}(1992)}]{palla1992}
{Palla}, F. \& {Stahler}, S.~W. 1992, \apj, 392, 667

\bibitem[{{Peters} {et~al.}(2010){Peters}, {Banerjee}, {Klessen}, {Mac Low},
  {Galv{\'a}n-Madrid}, \& {Keto}}]{peters2010a}
{Peters}, T., {Banerjee}, R., {Klessen}, R.~S., {et~al.} 2010, \apj, 711, 1017

\bibitem[{{Shu}(1977)}]{shu1977}
{Shu}, F.~H. 1977, \apj, 214, 488

\bibitem[{{Stahler}(1983)}]{stahler1983}
{Stahler}, S.~W. 1983, \apj, 274, 822

\bibitem[{{Stahler} {et~al.}(1986{\natexlab{a}}){Stahler}, {Palla}, \&
  {Salpeter}}]{stahler1986b}
{Stahler}, S.~W., {Palla}, F., \& {Salpeter}, E.~E. 1986{\natexlab{a}}, \apj,
  308, 697

\bibitem[{{Stahler} {et~al.}(1986{\natexlab{b}}){Stahler}, {Palla}, \&
  {Salpeter}}]{stahler1986a}
{Stahler}, S.~W., {Palla}, F., \& {Salpeter}, E.~E. 1986{\natexlab{b}}, \apj,
  302, 590

\bibitem[{{Stahler} {et~al.}(1980{\natexlab{a}}){Stahler}, {Shu}, \&
  {Taam}}]{stahler1980a}
{Stahler}, S.~W., {Shu}, F.~H., \& {Taam}, R.~E. 1980{\natexlab{a}}, \apj, 241,
  637

\bibitem[{{Stahler} {et~al.}(1980{\natexlab{b}}){Stahler}, {Shu}, \&
  {Taam}}]{stahler1980b}
{Stahler}, S.~W., {Shu}, F.~H., \& {Taam}, R.~E. 1980{\natexlab{b}}, \apj, 242,
  226

\bibitem[{{Wolfire} \& {Cassinelli}(1987)}]{wolfire1987}
{Wolfire}, M.~G. \& {Cassinelli}, J.~P. 1987, \apj, 319, 850

\bibitem[{{Yorke} \& {Kruegel}(1977)}]{yorke1977}
{Yorke}, H.~W. \& {Kruegel}, E. 1977, \aap, 54, 183

\bibitem[{{Yorke} \& {Sonnhalter}(2002)}]{yorke2002}
{Yorke}, H.~W. \& {Sonnhalter}, C. 2002, \apj, 569, 846

\end{thebibliography}

\end{document}